\documentclass[reprint, superscriptaddress, secnumarabic,amssymb, nobibnotes, aps, prl]{revtex4-1}

\usepackage{graphicx}
\usepackage{epstopdf}
\usepackage[T1]{fontenc}
\usepackage[latin9]{inputenc}
\usepackage{amsbsy}
\usepackage{gensymb}
\usepackage[T1]{fontenc}
\usepackage[latin9]{inputenc}
\usepackage{amsmath}
\usepackage{amssymb}
\usepackage{bbm}
\usepackage{braket}
\usepackage{xcolor}
\allowdisplaybreaks
\usepackage{graphicx}
\usepackage[colorlinks=true]{hyperref}
\usepackage{orcidlink}
\hypersetup{
    bookmarks=true,         
    unicode=false,          
    pdftoolbar=true,        
    pdfmenubar=true,        
    pdffitwindow=false,     
    pdfstartview={FitH},    
    pdftitle={Exploring the role of crystal structure in Rhenium-based superconductors},    
    pdfauthor={R. K. Kushwaha, R. P. Singh},     
    pdfsubject={},   
    pdfcreator={},   
    pdfproducer={}, 
    pdfkeywords={} {} {}, 
    pdfnewwindow=true,      
    colorlinks=true,       
    linkcolor=blue, 
    citecolor=blue,        
    filecolor=magenta,      
    urlcolor=blue           
}
\usepackage[normalem]{ulem}

\newcommand{\equref}[1]{Eq.~(\ref{#1})}

\newcommand{\figref}[1]{Fig.~\ref{#1}}

\newcommand{\tableref}[1]{Table~\ref{#1}}

\renewcommand{\approx}{\simeq}

\begin{document}
\title{\textrm{Superconductivity in new family of Rhenium-based binary alloys: Re$_{7}$X$_{3}$ (X = Nb, Ta, Ti, Zr, Hf)}}
\author{{R.~K.~Kushwaha}\,\orcidlink{0009-0005-3457-3653}}
\affiliation{Department of Physics, Indian Institute of Science Education and Research Bhopal, Bhopal, 462066, India}
\author{P. K. Meena}
\affiliation{Department of Physics, Indian Institute of Science Education and Research Bhopal, Bhopal, 462066, India}
\author{S. Jangid}
\affiliation{Department of Physics, Indian Institute of Science Education and Research Bhopal, Bhopal, 462066, India}
\author{P. Manna}
\affiliation{Department of Physics, Indian Institute of Science Education and Research Bhopal, Bhopal, 462066, India}
\author{S. Srivastava}
\affiliation{Department of Physics, Indian Institute of Science Education and Research Bhopal, Bhopal, 462066, India}
\author{T. Kumari}
\affiliation{Department of Physics, Indian Institute of Science Education and Research Bhopal, Bhopal, 462066, India}
\author{S. Sharma}
\affiliation{Department of Physics, Indian Institute of Science Education and Research Bhopal, Bhopal, 462066, India}
\author{P. Mishra}
\affiliation{Department of Physics, Indian Institute of Science Education and Research Bhopal, Bhopal, 462066, India}
\author{{R.~P.~Singh}\,\orcidlink{0000-0003-2548-231X}}
\email[]{rpsingh@iiserb.ac.in} 
\affiliation{Department of Physics, Indian Institute of Science Education and Research Bhopal, Bhopal, 462066, India}

\begin{abstract}
\begin{flushleft}
\end{flushleft}
Rhenium-based superconductors have recently attracted significant interest due to their unconventional superconducting properties. In this work, we report the synthesis and properties of new superconducting Re$_{7}$X$_{3}$ (X = Nb, Ta, Ti, Zr, Hf) binary alloys which maintain a fixed composition of rhenium while crystallizing in centrosymmetric to non-centrosymmetric crystal structures, depending on the elements of the X site. Comprehensive structural and superconducting properties were investigated using powder x-ray diffraction, AC transport, magnetization, and specific heat measurements, and on the basis of these measurements, the superconducting phase diagram was constructed. The results suggest a complex interplay of crystal structure and the Re/X ratio, which governs the strength of spin-orbital coupling and controls the unconventional superconducting behavior in Re-based superconductors.
\end{abstract}
\maketitle
\section{Introduction}
The unconventional superconductor pairing mechanism, along with its unique physical properties, offers a new paradigm for advanced understanding of the complex properties of quantum materials and their practical application in quantum technologies~\cite{H2S,BSCCO,YBCO,CePt3Si,CeCoIn5,UT3,UB3,orgSC,FeSC,p-wave1,p-wave2,topoSC}. Recently, Re-based superconductors have emerged as a new class of superconductors, where non-centrosymmetric (NCS) and centrosymmetric (CS) superconductors have recurrent occurrences of time-reversal symmetry breaking (TRSB)~\cite{TRSBreview}, which is a rarely observed phenomenon~\cite{Re6Zr,Re6Ti,Re6Hf,ReMoTRSB,Re2Hf,ReTRSB}. Despite decades of research, understanding the superconducting pairing mechanism and its unconventional properties in Re-based superconductors remains elusive. The presence of TRSB in the NCS crystal structure is mainly attributed to the presence of mixed spin-singlet and spin-triplet Cooper pairing states, as the lack of inversion symmetry allows an antisymmetric spin-orbit coupling (ASOC) that can lift the degeneracy of conduction band electrons~\cite{TRSBreview,ASOC}. At the same time, the superconducting presence of TRSB in Re- based CS superconductors is attributed to the critical percentage of elemental Re and its local microstructure~\cite{ReMoTRSB,ReTRSB}. However, the absence of such symmetry breaking in certain NCS superconductors~\cite{Sigrist,Smidman,Re3Ta,Re3W,Re5.5Ta,ReB} and its presence and absence in CS Re-based alloys~\cite{ReMoTRSB,Re2Hf} and elemental Re pose additional questions regarding the mechanism behind TRSB in Re-based superconductors~\cite{ReTRSB}.\\ 
To understand the unconventional superconducting properties and the pairing mechanism in Re-based superconductors, we have to understand the role of crystal structure, the strength of parity mixing, the elemental composition, the local microstructure, and the strength of spin-orbit coupling (SOC)~\cite{Sigrist,Smidman,TRSBreview}. However, a comprehensive understanding of these parameters has not yet been achieved. Currently, the primary focus of research on rhenium (Re)-based superconductors revolves around binary alloys involving transition metals (T), crystallized in CS or NCS crystal structures having fixed Re-concentration or Re-transition metal (T) solid solutions (Re$_{1-x}$T$_{x}$), where variations in composition at the Re and T sites result in crystallization in CS or NCS crystal structures~\cite{Re6Zr,Re6Ti,Re6Hf,Re-Mo,ReMoTRSB,ReNb,Re3Ta,Re3W,Re5.5Ta,Re2Hf,ReTRSB}. Binary alloys having fixed Re-atomic percentages and varying T sites with 3d, 4d, and 5d elements crystallizing in the CS and NCS crystal structures can be an ideal platform to study the role of crystal structure, ASOC strength, critical atomic percentage, and disorder in superconducting properties of Re-based superconductors. Re$_{7}$X$_{3}$ (X = Nb, Ta, Ti, Zr, Hf) is one of the superconducting series, which provides an excellent opportunity to explore the role of SOC by involving various 3d, 4d, and 5d elements at the X site, which covers both types of crystal structure$-$ symmorphic NCS and nonsymmorphic CS. Furthermore, Re$_{7}$X$_{3}$ series compounds provide an opportunity for comparative analysis with the hexagonal NCS compound Re$_{7}$B$_{3}$ which can offer additional information on the influence of crystal structure, elemental composition, and the role of the X site on the superconducting properties of Re-based superconductors.\\
In this work, we report the synthesis and comparative study of a new family of Re-based binary alloys Re$_{7}$X$_{3}$ (X = Nb, Ta, Ti, Zr, Hf), using magnetization, transport, and specific heat measurements. Based on these measurements, we have created superconducting phase diagrams to probe the complex interplay of different superconducting parameters, aiming to uncover the underlying nature of superconductivity in Re-based binary alloys.
\begin{figure*} [t!] 
\includegraphics[width=2.0\columnwidth,origin=b]{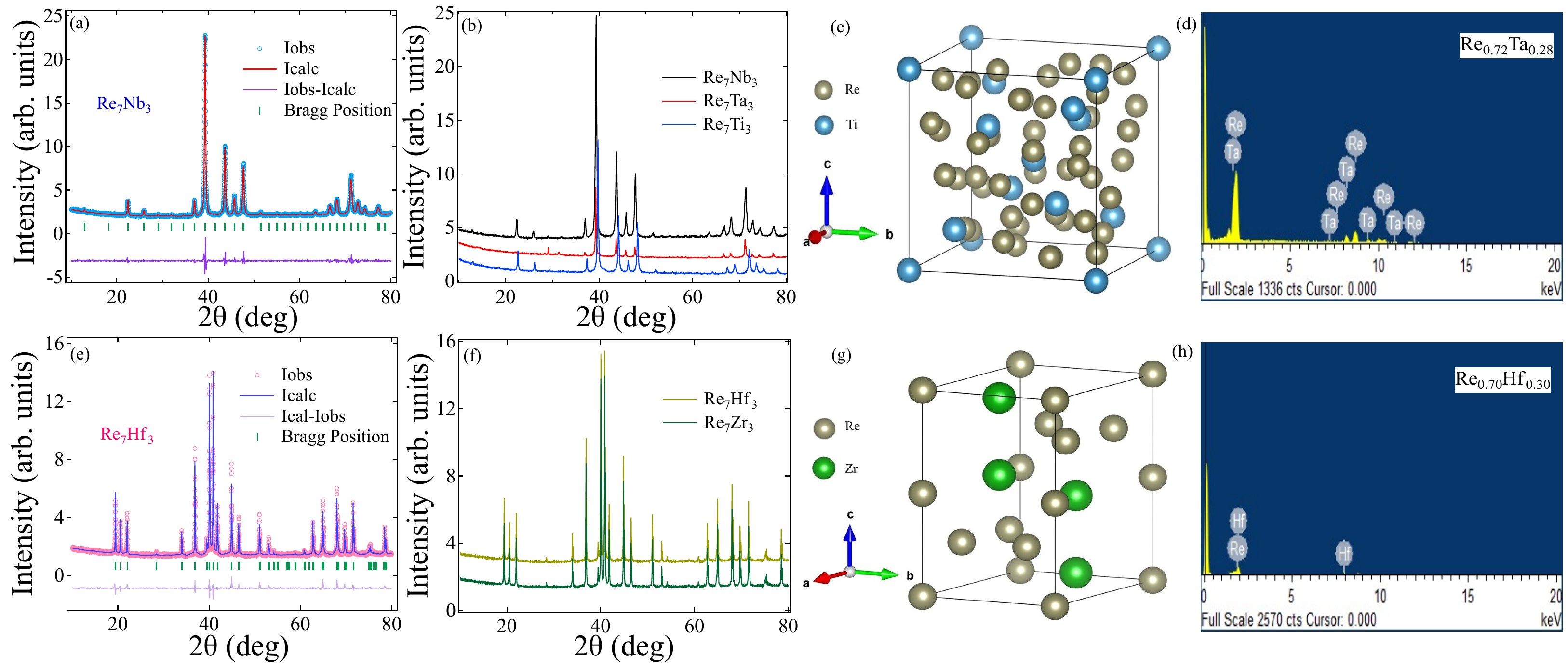}
\caption{\label{XRD}(a) and (e) Room-temperature refined powder XRD patterns for Re$_{7}$X$_{3}$ (X = Nb, Hf). (b) and (f) XRD patterns of NCS Re$_{7}$X$_{3}$ (X = Ti, Nb, Ta) and CS Re$_{7}$X$_{3}$ (X= Zr, Hf), respectively. (c) and (g) Crystal structure of NCS Re$_{7}$Ti$_{3}$ and CS Re$_{7}$Zr$_{3}$. (d) and (h) EDAX spectra of NCS Re$_{7}$Ta$_{3}$ and CS Re$_{7}$Hf$_{3}$.}
\end{figure*}

\section{Experimental Details}
Polycrystalline samples of Re$_{7}$X$_{3}$, (X = Nb, Ta, Ti, Zr, Hf) series were prepared by arc melting. High-purity metals (3N) Re, Nb, Ta, Ti, Hf and Zr in a stoichiometric ratio were melted several times to ensure the homogeneity of the samples in a pure argon atmosphere. Phase purity and crystal structure analysis were performed using a CuK$_{\alpha}$ (1.5406~\text{\AA}) PANalytical powder x-ray diffractometer. Fullprof software~\cite{Fullprof} was used to refine XRD patterns. Energy-dispersive x-ray (EDAX) analysis was performed using an Oxford scanning electron microscope (SEM) to observe the compositional information about the elements in our polycrystalline samples. Magnetization measurements were performed using a Quantum Design magnetic property measurement system (MPMS~3). Electrical resistivity and specific heat were measured on a Quantum Design Physical Property Measurement System (PPMS).

\section{Experimental Results}
\subsection{Sample Characterization}
Powder X-ray diffraction (XRD) patterns at room temperature for the five polycrystalline samples are shown in \figref{XRD} (a), (b), (e) and (f). Rietveld refinement confirms the phase purity and crystal structure of all of the compounds. Re$_{7}$Nb$_{3}$, Re$_{7}$Ta$_{3}$, and Re$_{7}$Ti$_{3}$ crystallise in a cubic ${\alpha}$-Mn structure with a symmorphic space group of I$\Bar{4}3m$. \figref{XRD}(a) and \figref{XRD}(b) show the Rietveld refinement of Re$_{7}$Nb$_{3}$ and the collective powder XRD patterns of all NCS phases. Conversely, Re$_{7}$Hf$_{3}$ and Re$_{7}$Zr$_{3}$ crystallize in the hexagonal centrosymmetric C14 Laves phase. The phase purity of Re$_{7}$Hf$_{3}$ is confirmed by the Rietveld refinement, shown in \figref{XRD}(e). \figref{XRD}(f) displays the powder XRD patterns of both CS Laves phases. Furthermore, \figref{XRD}(c) and (g) illustrate the crystal structures of NCS Re$_{7}$Ti$_{3}$ and CS Re$_{7}$Zr$_{3}$, respectively. Elemental composition analysis via EDAX confirms the stoichiometric ratio of Re$_{7}$X$_{3}$. EDAX spectra for NCS Re$_{7}$Ta$_{3}$ and CS Re$_{7}$Hf$_{3}$ are presented in \figref{XRD}(d) and (f), respectively. The lattice parameters, cell volume, and structure obtained from refinement are summarized in \tableref{Tb1}.
\begin{table}[h]
\caption{Structural parameters for Re$_{7}$X$_{3}$ (X = Nb, Ta, Ti, Hf, Zr) obtained from XRD refinement.}
\label{Tb1}
\begin{center}
\begingroup
\setlength{\tabcolsep}{3pt}
\begin{tabular}[t]{c c c c c c}\hline\hline
Parameters&Re$_{7}$Nb$_{3}$&Re$_{7}$Ta$_{3}$&Re$_{7}$Ti$_{3}$&Re$_{7}$Hf$_{3}$&Re$_{7}$Zr$_{3}$\\
\hline                                  
Structure& $\alpha$-Mn& $\alpha$-Mn& $\alpha$-Mn & Hexagonal& Hexagonal\\
Space group& $I$-43$m$&$I$-43$m$&$I$-43$m$& P63/${mmc}$& P63/${mmc}$\\
a = b (\AA)& 9.7181& 9.6216& 9.6223& 5.2445& 5.2665\\
c (\AA)& 9.7181& 9.6216& 9.6223& 8.5845& 8.6261\\
V$_{cell}$ (\AA$^{3}$)& 917.78& 890.71& 890.90& 204.47 &207.20\\
\hline\hline
\end{tabular}
\par\medskip\footnotesize
\endgroup
\end{center}
\end{table}

\subsection{Resistivity}
\begin{figure} [t] 
\includegraphics[width=0.98\columnwidth,origin=b]{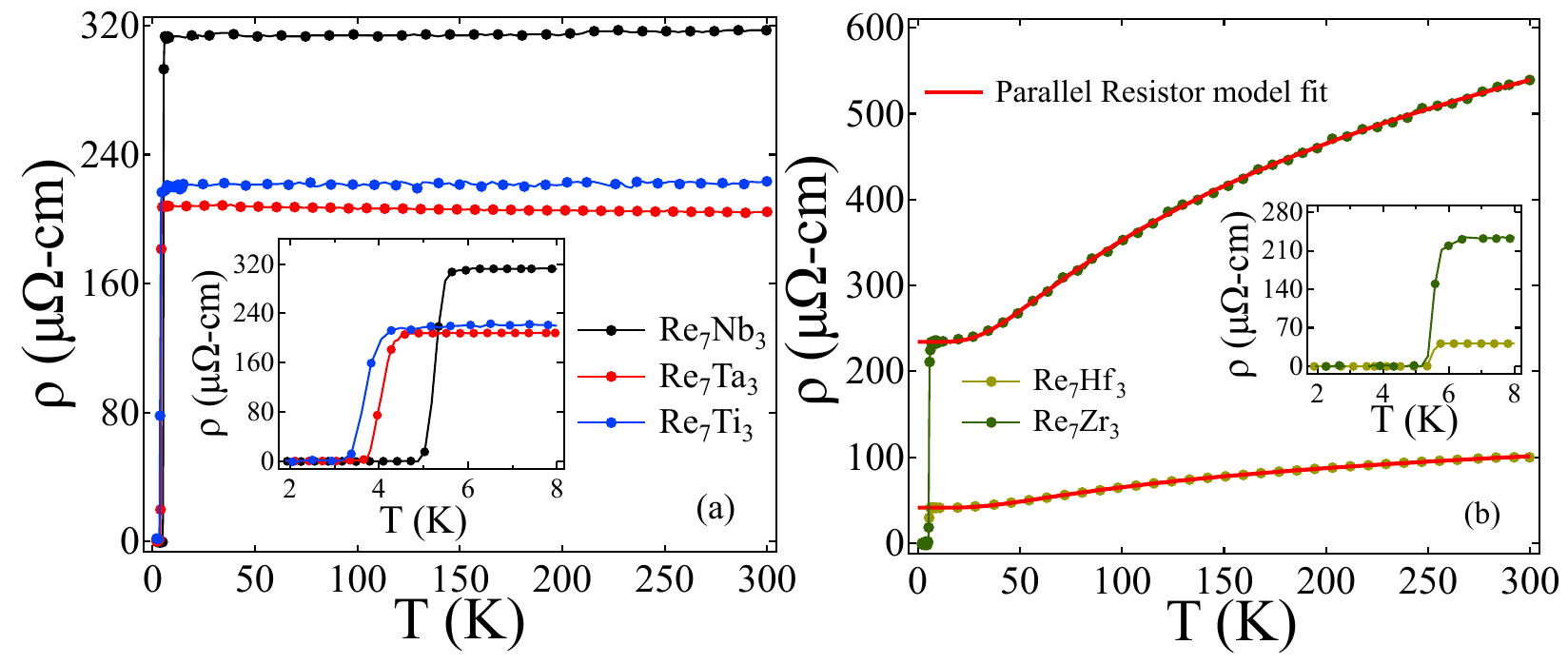}
\caption{\label{RT}Temperature-dependent zero-field resistivity of (a) Re$_{7}$X$_{3}$ (X = Nb, Ta, Ti) and (b) Re$_{7}$X$_{3}$ (X = Zr, Hf). Inset shows zero drops in resistivity near $T_{C}$ for Re$_{7}$X$_{3}$.}
\end{figure}

Temperature-dependent resistivity ($\rho(T)$) in the zero field confirmed the transition temperature $T_{C}$ = 5.2(2), 3.6(4), 3.8(3), 5.6(5), 6.0(5)~K for Re$_{7}$X$_{3}$ (X = Nb, Ta, Ti, Hf, Zr), respectively, as shown in \figref{RT}(a) and (b). Their insets show the zero drops in resistivity. The low value of the residual resistivity ratio (RRR) suggested a disordered nature of the NCS compounds; however, the CS phase in the RRR is better compared to the NCS compounds.
 $\rho(T)$ data in normal region up to room temperature, for Re$_{7}$Hf$_{3}$ and Re$_{7}$Zr$_{3}$ are well fitted to the parallel resistor model~\cite{parallel}. This model represents the expression for $\rho$(T) as \equref{para1}
\begin{equation}
\label{para1}
\frac{1}{\rho(T)} = \frac{1}{\overline{\rho}(T)}+\frac{1}{\rho_{sat}}
\end{equation}
where, $\rho_{sat}$ stands for saturation resistivity and the temperature-dependent resistivity $\overline{\rho}(T)$ represented by the following \equref{para2}
\begin{equation}
\label{para2}
\overline{\rho}(T)=\rho_{0}+b \left({\frac{T}{\theta_{D}}}\right)^{n}\int_{0}^{\theta_{D}/T}\frac {x^{n}}{(e^{x}-1)(1-e^{-x})}dx
\end{equation}
where $\rho_{0}$ is residual resistivity, $\theta_{D}$ is the Debye temperature, and b is a material-dependent factor. The second term in \equref{para2} contributes due to phonon-mediated electron scattering.
Utilizing \equref{para2} for $n$ = 5, \equref{para1} yields the best fit in the normal region for Re$_{7}$Hf$_{3}$ and Re$_{7}$Zr$_{3}$ as shown in \figref{RT}(b). The parameters obtained from the parallel resistor fittings are $\rho_{0}$ = 54.6(1) and 309.6(5)~$\mu\Omega\cdot cm$, $\rho_{sat}$ = 172.1(3) and 963.5(8)~$\mu\Omega\cdot cm$ and $\theta_{D}$ = 223.0(2) and 212.8(4)~K for Re$_{7}$Hf$_{3}$ and Re$_{7}$Zr$_{3}$, respectively, which are close to the $\theta_{D}$ values obtained from our zero-field specific heat measurements (see Section \ref{SHsection}).

\subsection{Magnetization}
\begin{figure} [t] 
\includegraphics[width=0.94\columnwidth,origin=b]{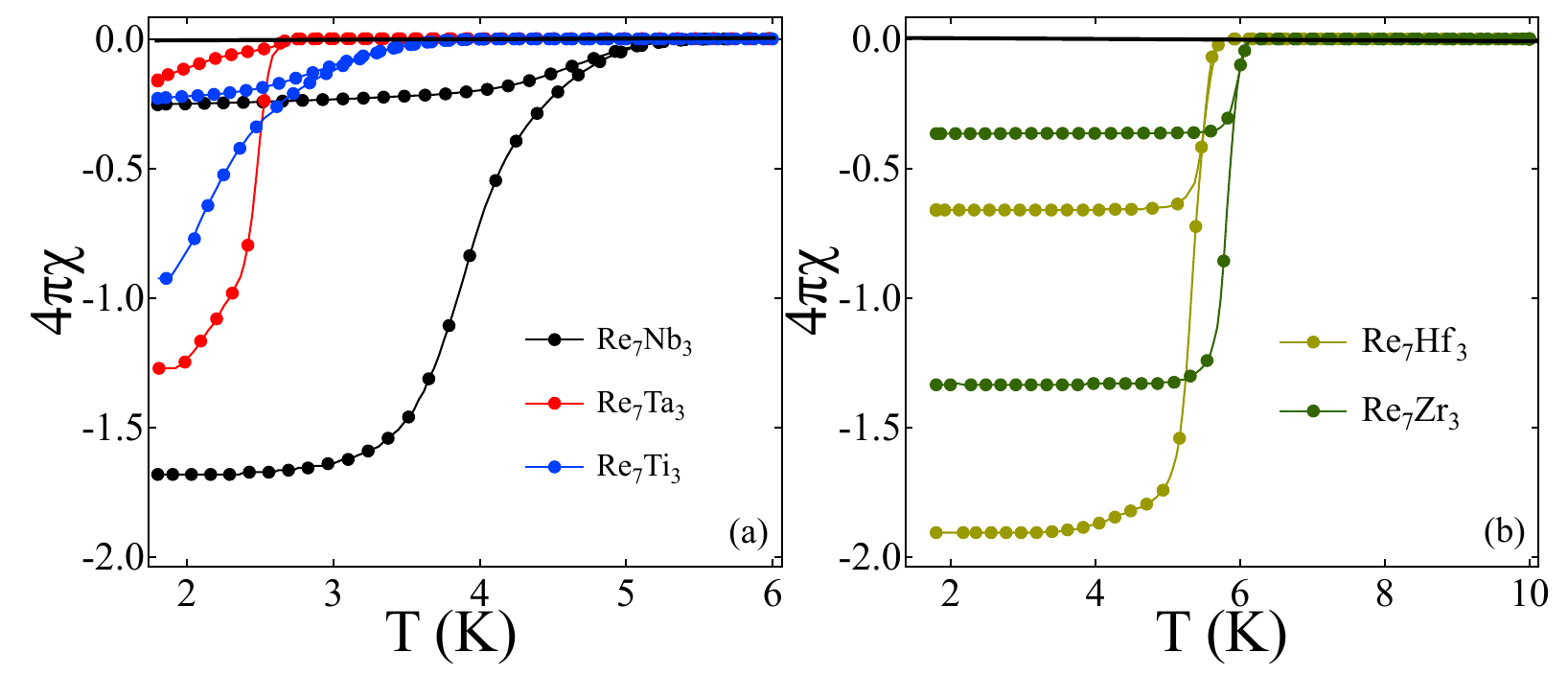}
\caption{\label{MT} Temperature-dependent magnetic susceptibility in ZFCW and FCC mode for (a) Re$_{7}$Nb$_{3}$, Re$_{7}$Ta$_{3}$, and Re$_{7}$Ti$_{3}$ (b) Re$_{7}$Hf$_{3}$ and Re$_{7}$Zr$_{3}$.}
\end{figure}

\begin{figure} [b] 
\includegraphics[width=0.94\columnwidth,origin=b]{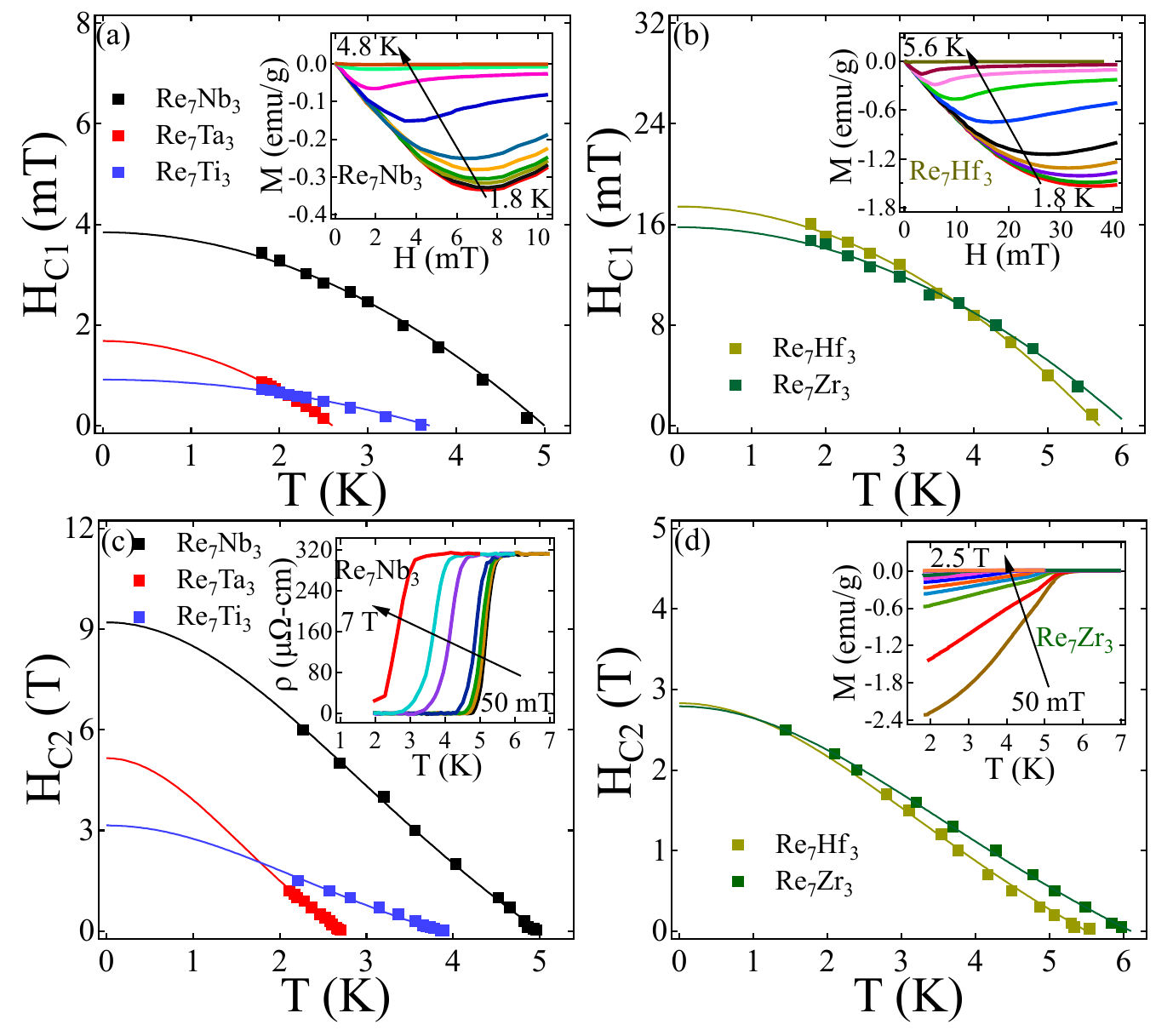}
\caption{\label{Hc}Temperature-dependent lower critical fields (a), (b) and upper critical field (c), (d) for Re$_{7}$X$_{3}$. Insets in (a) and (b) show magnetic field-dependent magnetization for Re${_7}$Nb${_3}$ and Re${_7}$Hf${_3}$ and (c) and (d) temperature-dependent resistivity for Re${_7}$Nb${_3}$ and temperature-dependent magnetization for Re${_7}$Zr${_3}$.}
\end{figure}

Temperature and field-dependent magnetization experiments were conducted on the five alloys using zero field-cooled warming (ZFCW) and field-cooled cooling (FCC) modes, with an applied field of 1~mT. All compounds exhibited bulk superconductivity. The onset temperature ($T_{C}$) of superconducting transitions for NCS and CS Re$_{7}$X$_{3}$ compounds is shown in \figref{MT}(a) and \figref{MT}(b), respectively.
The lower critical field, $H_{C1}(0)$, was determined from the low field-dependent magnetization curves observed at different temperatures. The lower critical field for each temperature curve was identified as the point where the curve diverges from linearity. A fit using the Ginzburg-Landau (GL) relation,
\begin{equation}
H_{C1}(T)=H_{C1}(0)\left[1-t^{2}\right],
\label{Hc1}
\end{equation}
where $t = T/T_{C}$, \equref{Hc1} yields $H_{C1}(0)$ = 3.91(1), 1.70(3), 0.95(7), 15.97(2), 17.4(1)~mT for Re$_{7}$X$_{3}$ (X = Nb, Ta, Ti, Zr, Hf), respectively. The variation of the lower critical field with temperature for NCS Re$_{7}$X$_{3}$, (X = Nb, Ta, Ti) have shown in \figref{Hc}(a) and CS Re$_{7}$X$_{3}$, (X = Zr, Hf)
 have shown in \figref{Hc}(b). The inset of both figures shows the variation of magnetization with the magnetic field at different temperatures for Re$_{7}$Nb$_{3}$ and Re$_{7}$Hf$_{3}$.

We have estimated the upper critical field $H_{C2}(0)$ using the field dependence of the superconducting onset temperature and the mid point of zero drop resistivity obtained from temperature-dependent magnetization and resistivity measurements. $H_{C2}(T)$ was plotted against temperature and can be well described by the GL relation (\equref{Hc2}, where $t = T/T_{C}$, the reduced-temperature) for the upper critical field.
\begin{equation}
\label{Hc2}
H_{C2}(T) = H_{C2}(0)\left[\frac{1-t^{2}}{1+t^{2}}\right]
\end{equation}
GL fits using \equref{Hc2} are shown in \figref{Hc}(c) and (d), yields $H_{C2}(0)$ =  9.20(7), 5.15(8), 3.08(9), 2.78(7), 2.88(3)~T for Re$_{7}$X$_{3}$ (X = Nb, Ta, Ti, Zr, Hf), respectively.
The orbital pair-breaking effect and Pauli limiting field effect are autonomous Cooper pair-breaking processes under the applied magnetic field. Orbital limiting field is described by the Werthamer-Helfand-Hohenberg (WHH)~\cite{WHH1, WHH2} model expressed by \equref{WHH}
\begin{equation}
H_{C2}^{orb}(0)= -\alpha T_{C} \left.\frac{dH_{C2}(T)}{dT}\right\vert_{T=T_{C}}
\label{WHH}
\end{equation}
\begin{figure} [t] 
\includegraphics[width=0.93\columnwidth,origin=b]{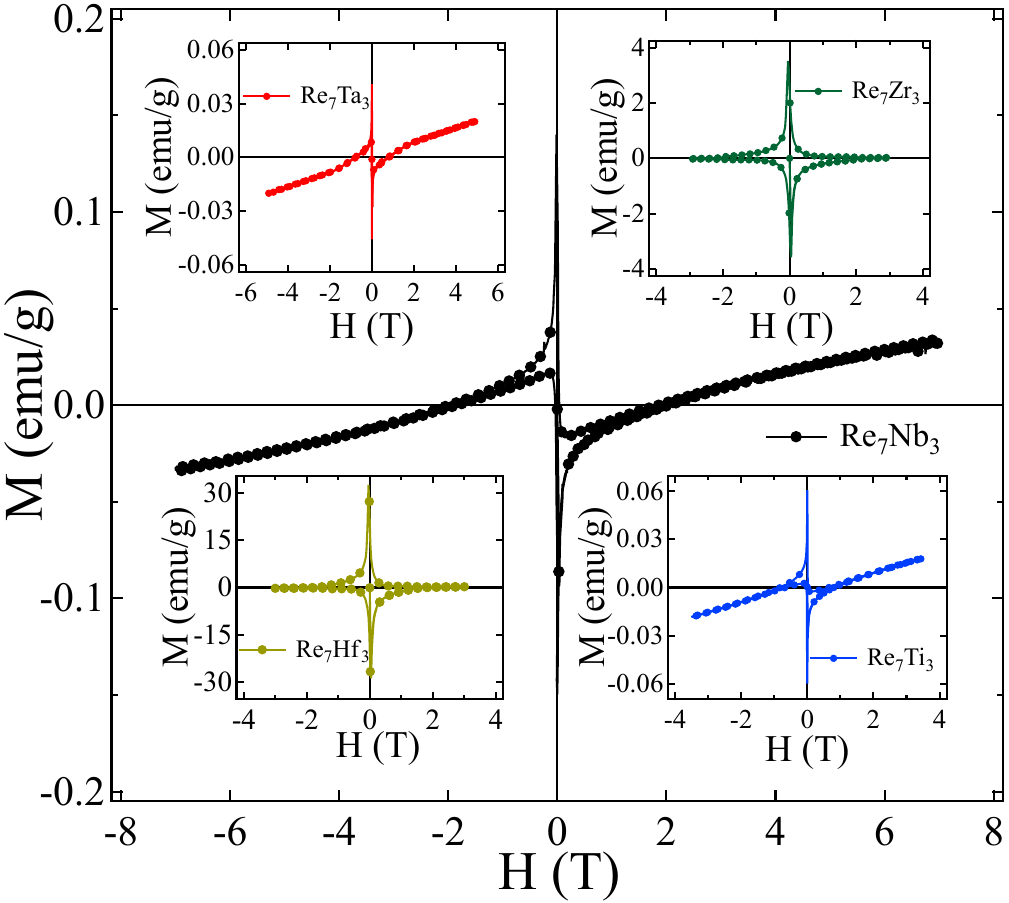}
\caption{\label{MHloop}M-H loop for Re$_{7}$Nb$_{3}$ and the insets show the M-H loops for Re$_{7}$X$_{3}$ (X = Ta, Hf, Ti, Zr) at 1.8~K.}
\end{figure}
Using the initial slopes $\frac{-dH_{C2}(T)}{dT}$ at $T = T_{C}$ for all samples and $\alpha$ = 0.693 in dirty limit superconductivity, the orbital limiting field was estimated as $H_{C2}^{orb}(0)$ = 7.25(1), 4.15(2), 1.01(5), 1.99(1), 1.13(8)~T for Re$_{7}$X$_{3}$ (X = Nb, Ta, Ti, Zr, Hf), respectively. For BCS superconductors, the Pauli limiting field $H_{C2}^{P} = 1.86 T_{C}$~\cite{Pauli1,Pauli2}. Using $T_{C}$ = 5.10, 2.69, 3.8, 6.1, 5.7~K, we have estimated $H_{C2}^{P}$ = 9.48(6), 5.00(3), 7.06(2), 11.3(4), 10.6(7)~T for Re$_{7}$X$_{3}$ (X = Nb, Ta, Ti, Zr, Hf), respectively. $H_{C2}$(0) in Re$_{7}$Nb$_{3}$ and Re$_{7}$Ta$_{3}$ close to the Pauli limiting field indicates the possibility of unconventional features in the superconducting state~\cite{CePt3Si,Re5.5Ta,Smidman}. Using Ginzburg-Landau (GL) theory, incorporating $H_{C1}(0)$ and $H_{C2}$(0) values, the coherence length $\xi_{GL}(0)$~\cite{Tinkham} is estimated from $H_{C2}(0)=\frac{\Phi_{0}}{2\pi\xi_{GL}^{2}}$ and penetration depth $\lambda_{GL}(0)$~\cite{lambda} is obtained from $H_{C1}(0)=\frac{\Phi_{0}}{4\pi\lambda_{GL}^2(0)}\left( ln \frac{\lambda_{GL}(0)}{\xi_{GL}(0)} + 0.12\right)$.
where, $\Phi_{0}$ = 2.07 $\times$10$^{-15}$ T m$^{2}$, the magnetic flux quantum~\cite{Tinkham}. Using the estimated values of $H_{C1}(0)$ and $H_{C2}(0)$, $\xi_{GL}(0)$ and $\lambda_{GL}(0)$ were evaluated to be 59.8(9), 80.0(2), 103.4(2), 109.2(6), 106.9(2)~$\text{\AA}$ and 4303(22), 6630(26), 8870(1460), 1724(49), 1643(21)~$\text{\AA}$ for Re$_{7}$X$_{3}$ (X = Nb, Ta, Ti, Zr, Hf), respectively. The GL parameter defined as $k_{GL}$ = $\frac{\lambda_{GL}(0)}{\xi_{GL}(0)}$ = 71.9(1), 82.9(1), 85.7(7), 15.8(4), 15.3(6) $\gg$ $\frac{1}{\sqrt{2}}$, indicating that all alloys in Re$_{7}$X$_{3}$ series are strong type-II superconductors. The thermodynamic critical field ($H_{C}(0)$) is defined as $H_{C1}(0)H_{C2}(0) = H_{C}(0)^2ln(k_{GL})$, using the estimated values of $H_{C1}(0)$ and $H_{C2}(0)$, we have obtained $H_{C}(0)$ = 91.7(2), 44.5(3), 25.7(2), 126.7(1), 135.4(3)~mT for Re$_{7}$X$_{3}$ (X = Nb, Ta, Ti, Zr, Hf), respectively. Magnetization vs field measurements at 1.8~K for all five alloys is shown in \figref{MHloop}. Irreversible fields ($H_{irr}$) were obtained as 2.23(4), 1.35(6), 1.27(1), 2.21(2), 2.27(6)~T for Re$_{7}$X$_{3}$ (X = Nb, Ta, Ti, Zr, Hf), respectively, above which the vortices unpinning begins. The Maki parameter $\alpha_{M} = \sqrt{2} \frac{H_{C2}^{orb}}{H_{C2}^{P}}$ is the relative strength of Pauli and the orbital limiting effect~\cite{maki}. $\alpha_{M}$ yields 1.08(1), 1.17(1), 0.20(2), 0.15(3), 0.25(2) for Re$_{7}$X$_{3}$, (X = Nb, Ta, Ti, Zr, Hf), respectively.

\subsection{Specific Heat}\label{SHsection}
\begin{figure} [t] 
\includegraphics[width=0.98\columnwidth,origin=b]{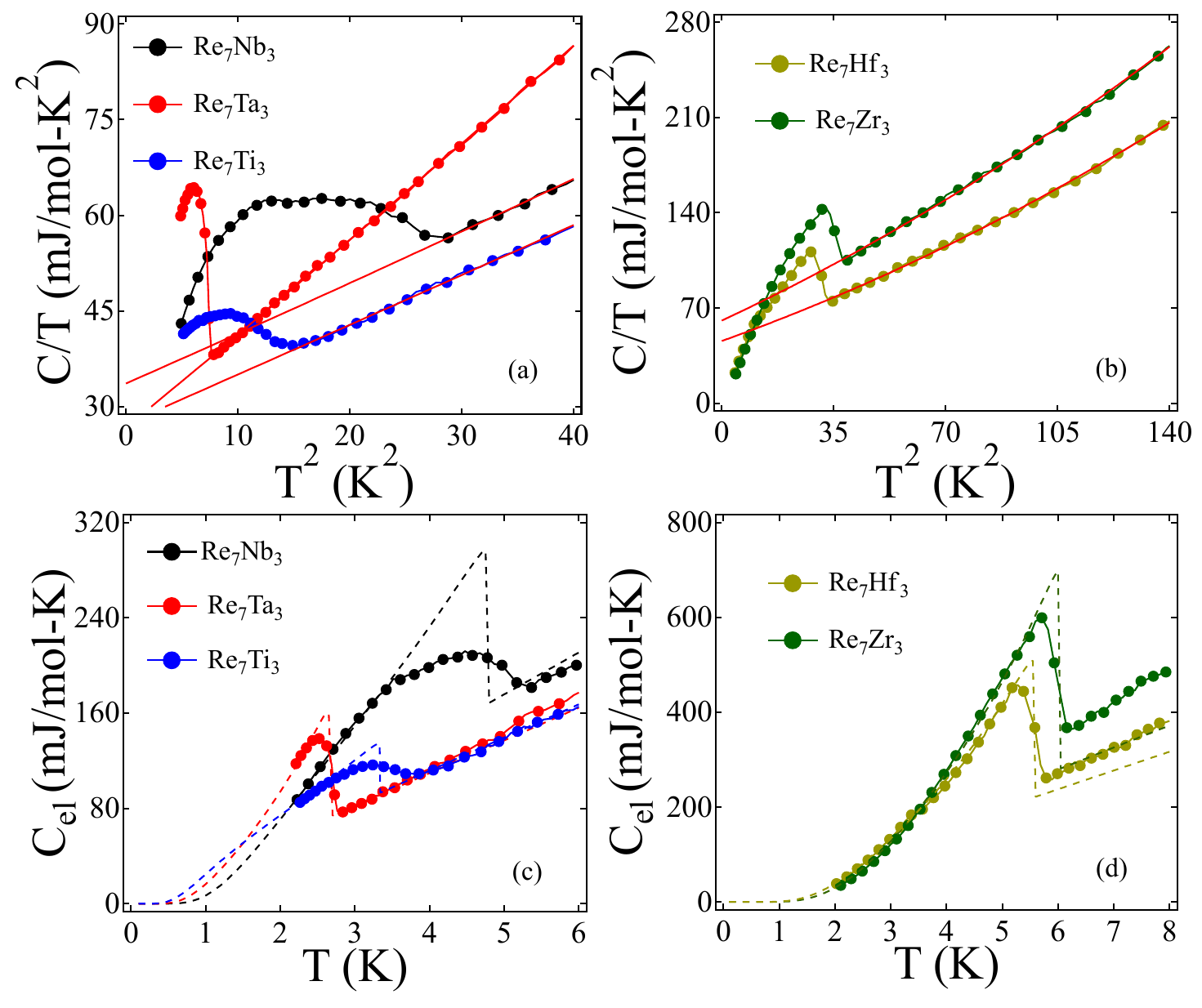}
\caption{\label{SH} Zero-field $C/T$ vs $T^{2}$ plots for (a) Re$_{7}$X$_{3}$, (X = Nb, Ta, Ti) and (b) Re$_{7}$X$_{3}$ (X = Zr, Hf), where the solid red lines represent the corresponding fit to \equref{C/T}. The temperature-dependent electronic specific heat for (c) Re$_{7}$X$_{3}$, (X = Nb, Ta, Ti) and (d) Re$_{7}$X$_{3}$ (X = Zr, Hf), fitted with the s-wave single gap model represented by the dotted lines.}
\end{figure}

Temperature-dependent specific heat measurements were performed on all samples in the zero field. $C/T$ vs $T^{2}$ data were fitted above $T_{C}$, using the Debye model represented by \equref{C/T}, as shown in \figref{SH}(a) and (b) with solid red curves.
\begin{equation}
C/T = \gamma_{n}+\beta_{3} T^{2}+\beta_{5} T^{4}
\label{C/T}
\end{equation}
where $\gamma_{n}$ is the Sommerfeld coefficient, $\beta_{3}$ and $\beta_{5}$ are the phononic and anharmonic contributions to the specific heat. Fitting normal state specific heat data using \equref{C/T} gives $\gamma_{n}$ = 33.9(2), 28.3(4), 27.2(5), 50.1(1), 46.6(2)~mJ mol$^{-1}$K$^{-2}$ and $\beta_{3}$ = 0.76(4), 1.38(2), 0.77(3), 1.40(2), 0.82(4)~mJ mol$^{-1}$K$^{-4}$ for Re$_{7}$X$_{3}$ (X = Nb, Ta, Ti, Zr, Hf), respectively. $\gamma_{n}$ is related to the density of states at Fermi level $D_{c}(E_{f})$ by the relation $\gamma_{n}$ = $\left(\frac{\pi^{2}k_{B}^{2}}{3}\right)D_{c}(E_{f})$, where $k_{B}$ $\approx$ 1.38 $\times$ 10$^{-23}$~J K$^{-1}$. $D_{c}(E_{f})$ is estimated to be 14.39(1), 12.02(1), 11.57(4), 21.25(2), 19.77(3)~states/eV f.u. for Re$_{7}$X$_{3}$ (X = Nb, Ta, Ti, Zr, Hf), respectively. Debye temperature ($\theta_{D}$) related to $\beta_{3}$ as $\theta_{D}$ = $\left(\frac{12\pi^{4}RN}{5\beta_{3}}\right)^{1/3}$, where $N$ is the number of atoms per formula unit and $R$ is the molar gas constant (8.314~J mol$^{-1}$ K$^{-1}$), gives $\theta_{D}$ = 293.9(8), 241.4(2), 292(1), 240(2), 287(4)~K for Re$_{7}$X$_{3}$ (X = Nb, Ta, Ti, Zr, Hf), respectively. It is similar to the $\theta_{D}$ values for other Re-based superconductors~\cite{Re5.5Ta,Re2Hf,Re3Ta}.\\
McMillan's model estimates the electron-phonon coupling strength from a dimensionless quantity $\lambda_{e-ph}$~\cite{McM}, which depends on the estimated values of $\theta_{D}$ and $T_{C}$ as
\begin{equation}
\lambda_{e-ph} = \left[\frac{1.04 + \mu^{*}ln(\theta_{D}/1.45T_{C})}{(1 - 0.62\mu^{*})ln(\theta_{D}/1.45T_{C}) - 1.04}\right]
\end{equation}
where $\mu^{*}$ is screened Coulomb repulsion, considering $\mu^{*}$ = 0.13 as described for the transition metals~\cite{McM}, we have obtained $\lambda_{e-ph}$ = 0.64(6), 0.57(2), 0.59(2), 0.73(3), 0.67(5) for Re$_{7}$X$_{3}$ (X = Nb, Ta, Ti, Zr, Hf), respectively. It classifies them as moderately strong coupled superconductors.
After subtracting phononic contributions from total specific heat, the temperature-dependent electronic specific heat for all samples was fitted with the isotropic s-wave gap model, represented in \figref{SH}(c) and (d). The quantity $\frac{\Delta C_{el}}{\gamma_{n}T_{C}}$ was found to be 0.94(4), 1.19(1), 0.46(5), 1.15(1), 1.08(2), for Re$_{7}$X$_{3}$ (X = Nb, Ta, Ti, Zr, Hf), respectively, which is less than the BCS limit (1.43).

Temperature-dependent electronic specific heat data can provide information about superconducting gap symmetry. Normalized entropy ($S$) in the superconducting region and the electronic-specific heat can be related as $C_{el} = t \frac{dS}{dt}$ where $t = T/T_{C}$, the reduced temperature. Within the framework of the BCS approximation~\cite{BCS1,BCS2}, normalized entropy for a single gap is defined by the following relation
\begin{equation}
\label{BCSeq}
\frac{S}{\gamma_{n}T_{C}} = -\frac{6}{\pi^2}\left(\Delta_{T}\right)\int_{0}^{\infty}[ \textit{f}\ln(f)+(1-f)\ln(1-f)]dy
\end{equation}
where, $\Delta_{T}=\frac{\Delta(0)}{k_{B}T_{C}}$, $\textit{f}$($\xi$) = [exp($\textit{E}$($\xi$)/$k_{B}T$)+1]$^{-1}$ is the Fermi function, $\textit{E}$($\xi$) = $\sqrt{\xi^{2}+\Delta^{2}(t)}$, where $E(\xi)$ is the energy of the normal electrons measured relative to the Fermi energy, $\textit{y}$ = $\xi/\Delta(0)$, and $\Delta(t)$ = tanh[1.82(1.018(($\mathit{1/t}$)-1))$^{0.51}$] resembles temperature-dependent superconducting energy gap~\cite{Re6Hf}. The dotted lines in \figref{SH}(c) and (d) represent the fit to electronic specific heat data using a single isotropic nodeless gap model as described by \equref{BCSeq}, which provides the normalized superconducting gap values $\frac{\Delta(0)}{k_B T_C}$ = 1.28(9), 1.63(2), 0.98(1), 1.84(2), 1.68(1) for Re$_{7}$X$_{3}$ (X = Nb, Ta, Ti, Zr, Hf), respectively. Re$_{7}$Zr$_{3}$ exhibits a gap value higher than the BCS limit (1.76), while other compounds display values below this limit, indicating weak electron-phonon coupling. The broadening in the specific heat jump in Re$_7$Nb$_3$ corresponds to the findings of a previous report on the NCS Nb$_{x}$Re$_{1-x}$ ~\cite{NbxRe1-x}.

\subsection{Elelctronic Properties and Uemura plot}

\begin{table*}
\caption{Superconducting and normal state parameters of Re$_{7}$X$_{3}$, (X = Nb, Ta, Ti, Zr, Hf).}
\label{tbl2}
\begin{center}
\begin{tabular}
{p{0.13\linewidth}p{0.13\linewidth}p{0.13\linewidth}p{0.13\linewidth}p{0.13\linewidth}p{0.13\linewidth}p{0.13\linewidth}}
\hline\hline
Parameters &unit &Re$_{7}$Nb$_{3}$ &Re$_{7}$Ta$_{3}$ &Re$_{7}$Ti$_{3}$ &Re$_{7}$Zr$_{3}$ &Re$_{7}$Hf$_{3}$\\
\hline
Structure& & NCS& NCS& NCS& CS& CS\\
$T_{C}$& K& 5.1(2)& 2.6(9)& 3.8(4)& 6.1(6)& 5.7(4)\\           
$H_{C1}(0)$& mT& 3.91(1)& 1.70(3)& 0.95(7)& 15.97(2)& 17.4(1)\\
$H_{C2}(0)$& T& 9.20(7)& 5.15(8) & 3.08(9)& 2.78(7)& 2.88(3)\\
$H_{C2}(0)^{Pauli}$& T& 9.48(6)& 5.00(3)& 7.06(2)& 11.3(4)& 10.6(7)\\
$H_{C2}(0)^{orb}$& T& 7.25(1)& 4.15(2)& 1.01(5)& 1.99(1)& 1.13(8) \\
$\xi_{GL}(0)$&  \text{\AA}& 59.8(9)& 80.0(2)& 103.4(2)& 109.2(6)& 106.9(3) \\
$\lambda_{GL}(0)$& \text{\AA}& 4303(22)& 6630(26) &8870(1460)& 1724(49)& 1643(21)\\
$k_{GL}$& &71.9(1) &82.9(1) &85.7(7)& 15.8(4)& 15.3(6)\\
$\Delta C_{el}/\gamma_{n}T_{C}$&   & 0.94(4)& 1.19(1)& 0.46(5)& 1.15(1)& 1.08(2)\\
$\Delta(0)/k_{B}T_{C}$&  &1.28(9)& 1.63(2)& 0.98(1)& 1.84(2)& 1.68(1)\\
$\theta_{D}$& K& 293.9(8)& 241.4(2) & 292(1)& 240(2)& 287(4)\\
$\lambda_{e-ph}$& & 0.64(6)& 0.57(2)& 0.59(2)&0.73(3)& 0.67(5)\\
$v_{F}$&$10^{4}$ m$s^{-1}$& 10.4& 11.06& 9.94& 9.02& 8.90\\
$n$&$10^{28}$ $m^{-3}$& 5.80& 4.84& 3.89& 8.59& 7.37\\
$T_{F}$&K& 4724& 4719& 3943& 4663& 4367\\
${T_{C}}/{T_{F}}$& & 0.001070& 0.000569& 0.000963& 0.001308& 0.001302\\
${m^{*}}/{m_{e}}$& & 13.31& 11.81& 12.22& 17.52& 16.89\\
\hline\hline
\end{tabular}
\end{center}
\par\medskip\footnotesize
\end{table*}

To investigate electronic properties, we have solved a set of equations simultaneously and derived the Uemura classification of superconductors~\cite{U1,U2} for Re$_{7}$X$_{3}$ (X = Nb, Ta, Ti, Zr, Hf). Sommerfeld coefficient ($\gamma_{n}$) is related to the quasiparticle number-density ($n$), defined as equation $\gamma_{n}= \left(\frac{\pi}{3}\right)^{2/3} \frac{k_{B}^{2} m^{\ast} V_{f.u.} n^{1/3}}{\hbar^{2} N_{A}}$ and the Fermi velocity ($v_{f}$) is related to the electronic mean free path ($l_{e}$) and n by the relations $l_{e} = \frac{3 \pi^{2} \hbar^{2}}{e^{2} \rho_{0} m\ast^{2} v_{F}^2}$ and $n=\frac{1}{3 \pi^{2}} \left(\frac{m^{\ast} v_{F}}{\hbar}\right)^{3}$, respectively, where $k_{B}$ is the Boltzmann constant, $m^{\ast}$ is the effective mass of the quasiparticles, $V_{f.u.}$ is the volume of the formula unit, $N_{A}$ is the Avogadro number and $\rho_{0}$ is the residual resistivity.\\
\begin{figure} [b] 
\includegraphics[width=0.95\columnwidth,origin=b]{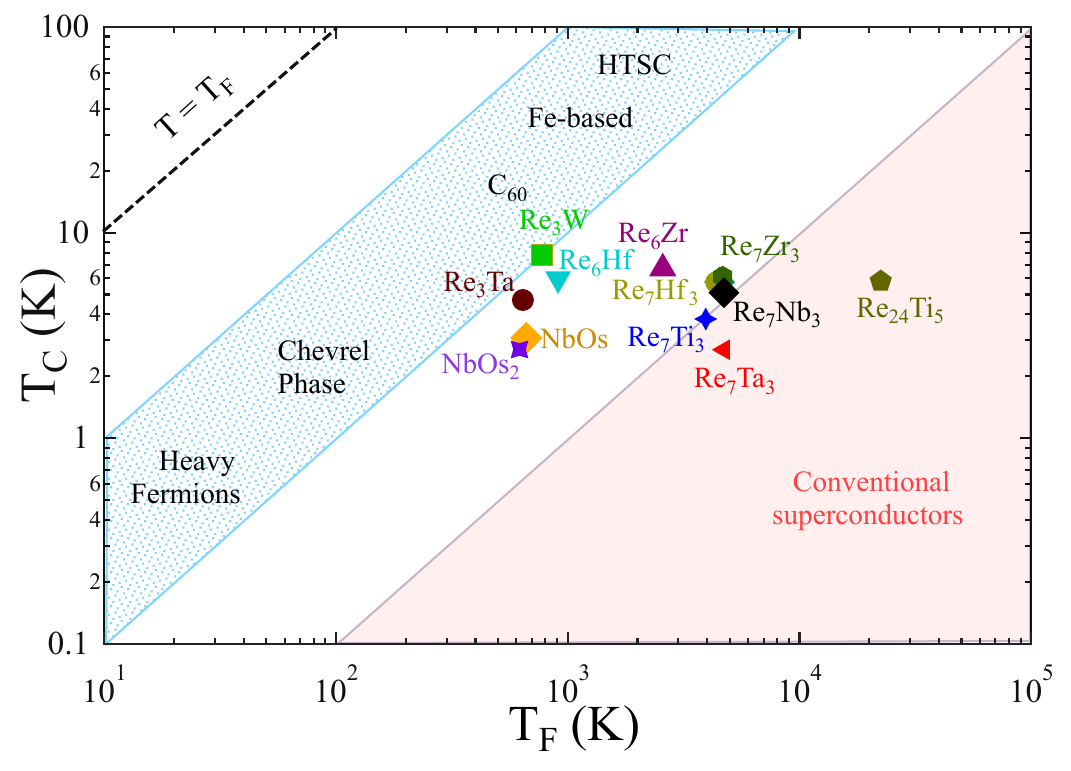}
\caption{\label{Uemura} Uemura plot for Re$_{7}$X$_{3}$ (X = Nb, Ta, Ti, Zr, Hf).}
\end{figure}
In dirty limit superconductivity, BCS coherence length is much larger than the mean free path ($\frac{\xi_{0}}{l_{e}} \gg 1$). As a result, the scattering of electrons with impurities may affect the superconducting properties. At T = 0~K, the GL penetration depth relates to the London penetration depth ($\lambda_{L}$) by expression $\lambda_{GL}(0)=\lambda_{L}(1+\frac{\xi_{0}}{l_{e}})^{1/2}$, where $\lambda_{L}=(\frac{m^{\ast}}{\mu_{0}ne^{2}})^{1/2}$ and BCS coherence length relates to G-L coherence length by the expression $\frac{\xi_{GL}(0)}{\xi_{0}}= \frac{\pi}{2\sqrt{3}}(1+\frac{\xi_{0}}{l_{e}})^{-1/2}$. As described in ref.~\cite{Re5.5Ta,Re3Ta}, we have solved the above equations simultaneously and estimated the values of $m^{\ast}$, $n$, $v_{F}$, $\xi_{0}$, $l_{e}$, using the values of $\gamma_{n}$, $\rho_{0}$, $\xi_{GL}(0)$ and $\lambda_{GL}(0)$.\\
The Fermi temperature for an isotropic spherical Fermi surface is defined as $T_{F}= \frac{\hbar^{2}}{2} \frac{k_{F}^{2/3}} {m^{\ast} k_{B}}$, where $k_{F}=3\pi^{2}n$, the Fermi wave vector. The ratio $T_{C}/T_{F}$ classifies superconductors into the conventional or unconventional category. According to Uemura et al.~\cite{U1,U2} the range 0.01$ \leq$ $\frac{T_{C}}{T_{F}}$ $\leq$ 0.1 is defined as an unconventional band. But $T_{C}/T_{F}$ for Re$_{7}$X$_{3}$ (X = Nb, Ta, Ti, Zr, Hf) lies outside this band near the other Re-based alloys, as shown in \figref{Uemura}. All superconducting and normal state parameters for Re$_{7}$X$_{3}$ (X = Nb, Ta, Ti, Zr, Hf) are summarized in \tableref{tbl2}.

\section{Superconducting Phase-diagram and Discussion}
\begin{figure} [b] 
\includegraphics[width=0.95\columnwidth,origin=b]{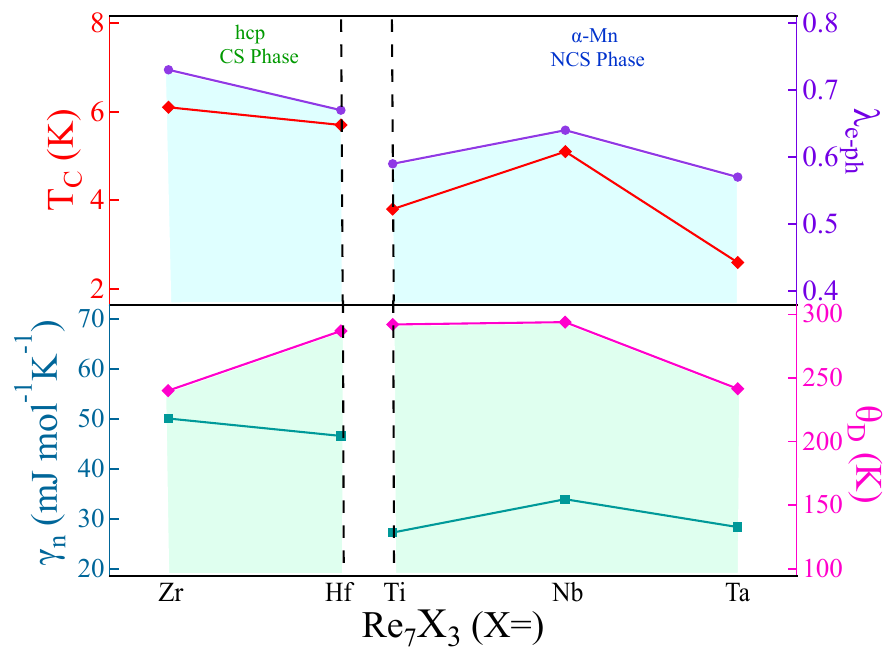}
\caption{\label{phase1} Superconducting phase diagram for Re$_{7}$X$_{3}$: upper section shows the variation in T$_{C}$ and $\lambda_{e-ph}$, while the lower section shows $\gamma_{n}$ and $\theta_{D}$ with respect to X. The left side denotes the CS phase, and the right side represents the NCS phase, separated by dotted lines.}
\end{figure}

\figref{phase1} shows the superconducting phase diagram for Re$_{7}$X$_{3}$ (X = Nb, Ta, Ti, Zr, Hf). This diagram delineates both the CS and NCS structures, distinguished by dotted lines. The upper section of \figref{phase1} illustrates the variations in T$_{C}$ and $\lambda_{e-ph}$ across different transition metals (X) within the Re$_{7}$X$_{3}$ series, indicating a decreasing trend in both parameters within the CS phase. This is correlated with reductions in the lattice parameters and cell volume. At the same time, the lower part of \figref{phase1} illustrates the behavior of the Debye temperature ($\theta_{D}$) and the Sommerfeld coefficient ($\gamma_{n}$), with $\theta_{D}$ increasing and $\gamma_{n}$ decreasing from Re$_{7}$Zr$_{3}$ to Re$_{7}$Hf$_{3}$.

\begin{figure} [t] 
\includegraphics[width=0.97\columnwidth,origin=b]{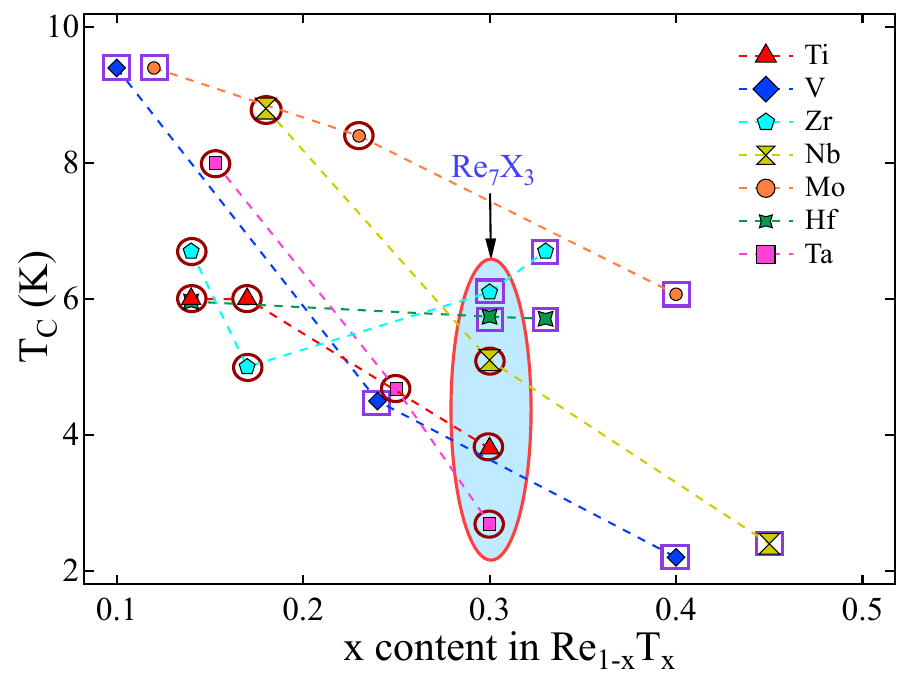}
\caption{\label{phase2}Superconducting phase diagram for Re$_{1-x}$T$_{x}$ (T = Ti, V, Zr, Nb, Mo, Hf, Ta)~\cite{Re6Ti,Re-V1,Re-V2,Re6Zr,Re-Zr,Re2Zr,ReNb,Re-Mo,Re6Hf,Re2Hf,Re3Ta,Re5.5Ta}, where purple squares represent CS phases and dark red circles represent NCS phases.}
\end{figure}

Interestingly, in the NCS phase, the lattice parameters and cell volume initially increase from Re$_{7}$Ti$_{3}$ to Re$_{7}$Nb$_{3}$, then decrease from Re$_{7}$Nb$_{3}$ to Re$_{7}$Ta$_{3}$, reflecting distinct trends in T$_{C}$, $\lambda_{e-ph}$, $\theta_{D}$, and $\gamma_{n}$. These trends are also consistent with variations in the $H_{C2}$ values and the lattice parameters, as shown in the upper and lower part of \figref{phase1} for the NCS region.\\
To further understand the influence of Re on superconducting properties, we have compiled composition and superconducting transition temperature data of Re$_{7}$X$_{3}$ and reported Re$_{1-x}$T$_{x}$ binary superconducting alloys~\cite{Re6Ti,Re-V1,Re-V2,Re6Zr,Re-Zr,Re2Zr,ReNb,Re-Mo,Re6Hf,Re2Hf,Re3Ta,Re5.5Ta}. \figref{phase2} illustrates the dependence of $T_{C}$ on the $x$ contents in Re$_{1-x}$T$_{x}$ alloys. Re$_{7}$X$_{3}$ (X = Nb, Ta, Ti, Zr, Hf) highlighted as a red oval. Purple squares denote CS phases, while dark red circles represent NCS phases. It is evident that the phases rich in Re consistently exhibit a decrease in $T_{C}$ as the percentage of Re decreases, except for the Re$_{1-x}$Zr$_{x}$.

Based on both above phase diagrams, we try to understand the interplay of crystal structure, elemental composition and SOC in the Re-based superconductor, which has been proposed as the key factor for unconventional superconductivity. Comparative study of cubic NCS superconductors Re$_{7}$X$_{3}$ and hexagonal NCS Re$_{7}$B$_{3}$ conventional superconductors~\cite{ReB}, both having the same Re compositions, provides a unique opportunity to unravel the superconducting properties of Re-based superconductors. Re-based, X site - 3d compound Re$_{7}$Ti$_{3}$ shows a reduced $H_{C2}$, $T_{C}$ and a superconducting gap compared to its isostructural alloys~\cite{Re6Ti}. However, with increasing Re concentration, compounds like Re$_{24}$Ti$_{5}$ and Re$_{6}$Ti exhibit $H_{C2}$ values comparable to the Pauli limit field, suggesting that the influence of the X site is negligible, underscoring the importance of the percentage of Re in structured compounds without centrosymmetry. Furthermore, when considering 4d- Zr and Nb and 5d- Hf and Ta, at the X site, Re$_{7}$Nb$_{3}$ and Re$_{7}$Ta$_{3}$ with NCS crystal structure have an upper critical field close to the Pauli limit. However, Re$_{7}$Hf$_{3}$ and Re$_{7}$Zr$_{3}$ with the CS crystal structure having the same Re concentration, the upper critical fields were much lower than the Pauli limit in emphasizing the role of non-centrosymmetry and ASOC in the unconventional superconducting properties of Re-based superconductors. This inference is further substantiated by comparing CS Re$_{3}$B and NCS Re$_{3}$(Ta, W)~\cite{Re3Ta,Re3W} and NCS Re$_{6}$X (X = Ti, Zr, Hf). These results suggest that the concurrent influence of crystal structure (CS/NCS), elemental percentage, and SOC strength could be the consolidated factors behind the origin of unconventional properties in Re-based superconductors.

\section{Conclusion}
In summary, we successfully synthesized a new series of Rhenium-based binary superconducting series of alloys, Re$_{7}$X$_{3}$, where X represents Nb, Ta, Ti, Zr and Hf, encompassing centrosymmetric (CS) and non-centrosymmetric (NCS) crystal structures. Our specific heat measurements have confirmed the presence of a nodeless gap with s-wave pairing across all samples, characterized by weak electron-phonon coupling. The upper critical field ($H_{C2}$) of the Re$_{7}$X$_{3}$ series with NCS structures closely approaches the Pauli limit, with the exception of Re$_7$Ti$_3$, while compounds with CS structures exhibit significantly lower $H_{C2}$ values compared to the Pauli limiting field. A comprehensive analysis of the compounds in the Re$_{7}$X$_{3}$ series with various Re-based superconductors underlines the intricate interplay between crystal structure, elemental composition, and antisymmetric spin-orbit coupling (ASOC) to determine superconducting properties. These findings suggest that the superconducting characteristics of Re-based superconductors can be fine-tuned by considering these factors carefully. Our results suggest that high-resolution crystallographic studies combined with density functional theory are necessary to elucidate the Fermi surface and ASOC effects and to comprehend unconventional superconducting phenomena like time-reversal symmetry breaking observed in both NCS and CS compounds, which have not been properly understood till date. Furthermore, the crystallization of Re$_{7}$X$_{3}$ compounds in intrinsically disordered NCS and less disordered non-symmorphic CS structures offers an ideal platform to explore the impact of the disorder on superconducting ground-state properties. Microscopic techniques such as $\mu$SR on the Re$_{7}$X$_{3}$ series compounds will further enhance our understanding of superconducting gap symmetry, the underlying pairing mechanism and the impact of the disorder on these characteristics. This, in turn, facilitates advances in understanding unconventional superconductivity in Re-based superconductors.

\section{Acknowledgement}
R.K.K. acknowledges the UGC, Government of India, for SRF fellowship. R.P.S. acknowledges the SERB Government of India for the Core Research Grant CRG/2023/000817.


\begin{thebibliography}{References}
\bibitem{H2S} A. P. Drozdov, M. I. Eremets, I. A. Troyan, V. Ksenofontov, and S. I. Shylin, Nature 525, 73 (2015).
\bibitem{BSCCO} H. Maeda, Y. Tanaka, M. Fukutomi, and T. Asano, Jpn. J. Appl. Phys. 27, L209 (1988).
\bibitem{YBCO} M. K. Wu, J. R. Ashburn, C. J. Torng, P. H. Hor, R. L. Meng, L. Gao, Z. J. Huang, Y. Q. Wang, and C. W. Chu, Phys. Rev. Lett. 58, 908 (1987).
\bibitem{CePt3Si} E. Bauer, G. Hilscher, H. Michor, Ch. Paul, E. W. Scheidt, A. Gribanov, Yu. Seropegin, H. Noël, M. Sigrist, and P. Rogl, Phys. Rev. Lett. 92, 027003 (2004).
\bibitem{CeCoIn5} C. Petrovic, P. G. Pagliuso, M. F. Hundley, R. Movshovich, J. L. Sarrao, J. D. Thompson, Z. Fisk, and P. Monthoux,  J. Phys. Condens. Matter 13, L337 (2001).
\bibitem{UT3} G. R. Stewart, Z. Fisk, J. O. Willis, and J. L. Smith, Phys. Rev. Lett. 52, 679 (1984).
\bibitem{UB3} H. R. Ott, H. Rudigier, Z. Fisk, and J. L. Smith, Phys. Rev. Lett. 50, 1595 (1983). 
\bibitem{orgSC} G. Saito, Y. Yoshida, The Chemical Record 11, 124 (2011).
\bibitem{FeSC} P. M. Aswathy, J. B. Anooja,  P. M. Sarun, and U. Syamaprasad, Supercond. Sci. Technol. 23, 073001 (2010).
\bibitem{p-wave1} A. P. Mackenzie, and Y. Maeno, Physica B: Condensed Matter 280, 1, (2000).
\bibitem{p-wave2} E. F.Talantsev, K. Iida, T. Ohmura, T. Matsumoto, W. P. Crump, N. M. Strickland, S. C. Wimbush, and H. Ikuta, Scientific Reports 9, 14245 (2019). 
\bibitem{topoSC} Y. Ando and L. Fu, Annu. Rev. Condens. Matter Phys. 6, 361 (2015).
\bibitem{TRSBreview} S. K. Ghosh, M. Smidman, T. Shang, J. F Annett, A. D. Hillier, J. Quintanilla and H. Yuan, J. Phys.: Condens. Matter 33, 033001 (2021).
\bibitem{Re6Zr} R. P. Singh, A. D. Hillier, B. Mazidian, J. Quintanilla, J. F. Annett, D. M. Paul, G. Balakrishnan, and M. R. Lees, Phys. Rev. Lett. 112, 107002 (2014).
\bibitem{Re6Hf} D. Singh, J. A. T. Barker, A. Thamizhavel, D. M. Paul, A. D. Hillier, and R. P. Singh, Phys. Rev. B 96, 180501 (2017).
\bibitem{Re6Ti} D. Singh, K. P. Sajilesh, J. A. T. Barker, D. M. Paul, A. D. Hillier, and R. P. Singh, Phys. Rev. B 97, 100505 (2018).
\bibitem{Re2Hf} M. Mandal, A. Kataria, C. Patra, D. Singh, P. K. Biswas, A. D. Hillier, T. Das, and R. P. Singh, Phys. Rev. B 105 (2022).
\bibitem{ReTRSB} T. Shang, M. Smidman, S. K. Ghosh, C. Baines, L. J. Chang, D .J. Gawryluk, J. A. T. Barker, R. P. Singh, D. McK. Paul, G. Balakrishnan, E. Pomjakushina, M. Shi, M. Medarde, A. D. Hillier, H. Q. Yuan, J. Quintanilla, J. Mesot, and T. Shiroka, Phys. Rev. Lett. 121, 257002 (2018).
\bibitem{ReMoTRSB} T. Shang, C. Baines, L. Chang, D. J. Gawryluk, E. Pomjakushina, M. Shi, M. Medarde and T. Shiroka, npj Quantum Materials 5, 76 (2020).
\bibitem{ASOC} L. P. Gor'kov and E. I. Rashba, Phys. Rev. Lett. 87, 037004 (2001).
\bibitem{Sigrist} E. Bauer and M. Sigrist, \textit{Non-centrosymmetric Superconductors}, Springer-Verlag, Berlin (2012).
\bibitem{Smidman} M. Smidman, M. B. Salamon, H. Q. Yuan, and D. F. Agterberg, Rep. Prog. Phys. 80, 036501 (2017).
\bibitem{Re5.5Ta} Arushi, D. Singh, P. K. Biswas, A. D. Hillier, and R. P. Singh, Phys. Rev. B 101, 144508 (2020).
\bibitem{Re3Ta}J. A. T. Barker, B. D. Breen, R. Hanson, A. D. Hillier, M. R. Lees, G. Balakrishnan, D. McK. Paul, and R. P. Singh, Phys. Rev. B 98, 104506 (2018).
\bibitem{ReB} S. Sharma, Arushi, K. Motla, J. Beare, M. Nugent, M. Pula, T. J. Munsie, A. D. Hillier, R. P. Singh, and G. M. Luke, Phys. Rev. B 103, 104507 (2021).
\bibitem{Re3W} P. K. Biswas, A. D. Hillier, M. R. Lees, and D. McK. Paul, Phys. Rev. B 85, 134505 (2012).
\bibitem{Re-Zr} K. Matano, R. Yatagai, S. Maeda, and Guo-qing Zheng, Phys. Rev. B 94, 214513 (2016).
\bibitem{Re2Zr} A. L. Giorgi and E. G. Szklarz, J. Less-Common Met. 22, 246 (1970).
\bibitem{Re-V1} B. W. Roberts, J. Phys. Chem. Ref. Data 5, 581 (1976).
\bibitem{Re-V2}J. L. Jorda and J. Muller, J. Less-Common Met 119, 337 (1986).
\bibitem{Re-Mo} T. Shang, D. J. Gawryluk, J. A. T. Verezhak, E. Pomjakushina, M. Shi, M. Medarde, J. Mesot, and T. Shiroka, Phys. Rev. Materials 3, 024801 (2019).
\bibitem{ReNb} A. B. Karki, Y. M. Xiong, N. Haldolaarachchige, S. Stadler, I. Vekhter, P. W. Adams, D. P. Young, W. A. Phelan, and J. Y. Chan, Phys. Rev. B 83, 144525 (2011).
\bibitem{Fullprof} J. Rodríguez-Carvajal, Phys. B: Cond. Matt. 192, 55 (1993).
\bibitem{parallel} H. Wiesmann, M. Gurvitch, H. Lutz, A. K. Ghosh, B. Schwarz, M. Strongin, P. B. Allen, and J. W. Halley, Phys. Rev. Lett. 38, 782 (1977).
\bibitem{WHH1} E. Helfand, and N. R. Werthamer, Phys. Rev. 147, 288 (1966).
\bibitem{WHH2} N. R. Werthamer, E. Helfand, and P. C. Hohenberg, Phys. Rev. 147, 295 (1966).
\bibitem{Pauli1} A. B. Karki, Y. M. Xiong, I. Vekhter, D. Browne, P. W. Adams, D. P. Young, K. R. Thomas, Julia Y. Chan, H. Kim, and R. Prozorov, Phys. Rev. B 82, 064512 (2010).
\bibitem{Pauli2} J. K. Bao, J. Y. Liu, C. W. Ma, Z. H. Meng, Z. T. Tang, Y. L. Sun, H. F. Zhai, H. Jiang, H. Bai, C. M. Feng, Z. A. Xu, and G. H. Cao, Phys. Rev. X 5, 011013 (2015).
\bibitem{Tinkham} M. Tinkham, \textit{Introduction to Superconductivity}, 2nd ed., McGraw-Hill, New York (1996).
\bibitem{lambda} T. Klimczuk, F. Ronning, V. Sidorov, R. J. Cava, and J. D. Thompson, Phys. Rev. Lett. 99, 257004 (2007).
\bibitem{maki} K. Maki, Phys. Rev. 148, 362 (1966).
\bibitem{McM} W. L. McMillan, Phys. Rev. 167, 331 (1968).
\bibitem{BCS1} J. Bardeen, L. N. Cooper, and J. R. Schrieffer, Phys. Rev. 108, 1175 (1957).
\bibitem{BCS2} H. Suhl, B. T. Matthias, and L. R. Walker, Phys. Rev. Lett. 3, 552 (1959).
\bibitem{NbxRe1-x} J. Chen, L. Jiao, J. L. Zhang, Y. Chen, L. Yang, M. Nicklas, F. Steglich, and H. Q. Yuan, Phys. Rev. B 88, 144510 (2013).
\bibitem{U1} Y. J. Uemura, V. J. Emery, A. R. Moodenbaugh, M. Suenaga, D. C. Johnston, A. J. Jacobson, J. T. Lewandowski, J. H. Brewer, R. F. Kiefl, S. R. Kreitzman, G. M. Luke, T. Riseman, C. E. Stronach, W. J. Kossler, J. R. Kempton, X. H. Yu, D. Opie, and H. E. Schone, Phys. Rev. B 38, 909(R) (1988).
\bibitem{U2} Y. J. Uemura, L. P. Le, G. M. Luke, B. J. Sternlieb, W. D. Wu, J. H. Brewer, T. M. Riseman, C. L. Seaman, M. B. Maple, M. Ishikawa, D. G. Hinks, J. D. Jorgensen, G. Saito, and H. Yamochi, Phys. Rev. Lett. 66, 2665 (1991).
\end{thebibliography}
\end{document}